# ISOSPIN VIOLATION IN HYPERON SEMILEPTONIC DECAYS

GABRIEL KARL

Department of Physics

University of Guelph

Guelph, ON, Canada N1G 2W1

Abstract: This note emphasizes that because of isospin violation, the two central states in the octet ($\Sigma°$, $\Lambda°$) mix, and this mixing can be measured in semileptonic decays, in particular with an accurate determination of $\Sigma^+$ semileptonic branching ratio.

The work I am describing has been published [1] some five years ago, but due to circumstances has not been advertized at conferences. The issue is isospin violation in matrix elements for semileptonic decays. At first sight this proposal looks hopeless since one might argue that $SU_3$ violation should be much more important. However, one may envisage scenarios where the $SU_3$ wavefunctions remain pure octet while isospin mixing inside the octet takes place due to the mass difference between up and down quarks. In any case, a number of authors have argued that there is a small mixing between $\Sigma_8^o$ and $\Lambda_8$ due to this mass difference [2], so that the physical states $\Sigma°, \Lambda$ are linear combinations of the pure isovector $\Sigma_8^o$ and the pure isoscalar $\Lambda_8$:

$$\Lambda = \Lambda_8 \cos\phi + \Sigma_8^o \sin\phi$$

$$\Sigma^o = -\Lambda_8 \sin\phi + \Sigma_8^o \cos\phi$$

with a mixing angle $\phi$:

$$\sin\phi = \phi = -\frac{\sqrt{3}}{4} \cdot \frac{m_d - m_u}{m_s - \hat{m}} \simeq -0.015$$

where $\hat{m}$ is the average mass of the up and down quarks, and $m_s$ is that of the strange one. These formulae are obtained both in the quark model and in the chiral quark model. The main point I wish to stress is that these are theoretical formulae which should be tested experimentally. The simplest direct tests are in semileptonic decays of hyperons which have a $\Lambda$ or $\Sigma^0$ in the initial or final state.

One such test involves the semileptonic decay of a charged $\Sigma$, say $\Sigma^-$ to a $\Lambda$. If isospin is conserved, the vector coupling vanishes [3], but with the mixing taken into consideration one finds [1,4]:

$$\left(g_V/g_A\right)_{\Sigma^- \to \Lambda} = \frac{\sqrt{3}}{D}\tan\phi = \frac{\sqrt{3}}{D}\phi \simeq -0.03$$

There is some data from CERN [5] which disagrees with the sign of this prediction, but the disagreement is not statistically significant. A precise measurement of $g_V$ in $\Sigma^-$ semileptonic decay would determine the magnitude of $\phi$.

An easier measurement [1,6] is the ratio of semileptonic decay rates for $\Sigma^+$ and $\Sigma^-$ where

$$R(\phi) = \frac{\Gamma\left(\Sigma^+ \to \Lambda \bar{e}\nu\right)}{\Gamma\left(\Sigma^- \to \Lambda e\bar{\nu}\right)} = R(0)\left(1 - \frac{4\sqrt{3}F}{D}\phi\right) \simeq 0.65$$

where $R(0)$ is taken from the review [7]. Although the experimental ratio agrees with 0.65, the agreement is not statistically significant. One needs to measure the semileptonic branching ratio for $\Sigma^+$, which is based at present on 21 events.

There are other, smaller corrections which are given in reference [1]. In principle, when analyzing the semileptonic data one should keep as parameters the Cabibbo angle $\theta$, the matrix elements F,D and the mixing angle ϕ to obtain a better determination of all these quantities.

The author was encouraged by learning at the Symposium that Dr. V. Smith is actively promoting these experiments.

The author is grateful to the organizers of Hyperon 99 for the opportunity to advertize these ideas, and to NSERC Canada for financial support.

References

[1]   Gabriel Karl, Physics Letters B$\underline{328}$, 149 (1994).

      Erratum: Phys. Let. B$\underline{341}$, 449 (1995).

[2]   N. Isgur, Phys. Rev. D$\underline{21}$, 779 (1980) and references therein.

      J. Gasser and H. Leutwyler, Phys. Rep. $\underline{87}$, 77 (1982).

      J.F. Donoghue, Ann. Rev. Nucl. Part. Sci. 39 (1989).

[3]   N. Cabibbo and R. Gatto, Nuovo Cimento $\underline{15}$, 159 (1960).

      V.P. Belov, B.S. Mingalev and V.M. Shekter, Sov. Phys. JET4$\underline{11}$, 392 (1960).

      N. Cabibbo and P. Franzini, Phys. Lett. $\underline{3}$, 217 (1963).

[4]   P. Bracken, A. Frenkel and G. Karl, Phys. Rev. D$\underline{24}$, 2984 (1984).

[5]   M. Bourquin et al., Z. Phys. C$\underline{12}$, 307 (1982).

[6]   E. Henley and J.E. Miller, Phys. Rev. D$\underline{50}$, 7077 (1994).

[7]   J.-M. Gaillard and G. Sauvage, Ann. Rev. Nucl. Part. Sci. $\underline{34}$, 351 (1984).